\documentclass{article}

\usepackage{PRIMEarxiv}

\usepackage[utf8]{inputenc} 
\usepackage[T1]{fontenc}    
\usepackage{hyperref}       
\usepackage{url}            
\usepackage{booktabs}       
\usepackage{amsfonts}       
\usepackage{amsmath}        
\usepackage{nicefrac}       
\usepackage{microtype}      
\usepackage{lipsum}
\usepackage{fancyhdr}       
\usepackage{graphicx}       
\graphicspath{{media/}}     
\usepackage{float}
\usepackage{moreverb}
\usepackage{mathtools}
\usepackage{setspace}

\title{Rematching on-the-fly: sequential matched randomization and a case for covariate-adjusted randomization}
  
\author{
  Jonathan J. Chipman \\
  Department of Population Health Sciences, Division of Biostatistics \\
  Cancer Biostatistics, Huntsman Cancer Institute \\
  University of Utah \\
  Salt Lake City, Utah\\
  \texttt{jonathan.chipman@hci.utah.edu} \\
   \And
  Lindsay Mayberry \\
  Department of Medicine, Vanderbilt University Medical Center \\
  Nashville, TN \\
  \AND
  Robert A. Greevy, Jr. \\
  Department of Biostatistics, Vanderbilt University Medical Center \\
  Nashville, TN \\
}

\begin{document}
\maketitle

\begin{abstract}
Covariate-adjusted randomization (CAR) can reduce the risk of covariate imbalance and, when accounted for in analysis, increase the power of a trial. Despite CAR advances, stratified randomization remains the most common CAR method. Matched Randomization (MR) randomizes treatment assignment within optimally identified matched pairs based on covariates and a distance matrix. When participants enroll sequentially, Sequentially Matched Randomization (SMR) randomizes within matches found “on-the-fly” to meet a pre-specified matching threshold. However, pre-specifying the ideal threshold can be challenging and SMR yields less-optimal matches than MR. We extend SMR to allow multiple participants to be randomized simultaneously, to use a dynamic threshold, and to allow matches to break and rematch if a better match later enrolls (Sequential Rematched Randomization; SRR). In simplified settings and a real-world application, we assess whether these extensions improve covariate balance, estimator/study efficiency, and optimality of matches. We investigate whether adjusting for more covariates can be detrimental upon covariate balance and efficiency as is the case of traditional stratified randomization. As secondary objectives, we use the case study to assess how SMR schemes compare side-by-side with common and related CAR schemes and whether adjusting for covariates in the design can be as powerful as adjusting for covariates in a parametric model. We find each SMR extension, individually and collectively, to improve covariate balance, estimator efficiency, study power, and quality of matches. We provide a case-study where CAR schemes with randomization-based inference can be as and more powerful than Non-CAR schemes with parametric adjustment for covariates.
\end{abstract}


\maketitle


\section{Introduction}\label{sec1}

Clinical trialists use covariate-adjusted randomization (CAR) schemes to reduce the risk of predictive covariate imbalances between treatment arms and to potentially increase the efficiency of a randomized trial\cite{rosenberger2015randomization}.  When an imbalance occurs, the trialist bears additional burden to convince the scientific community that an observed effect is not confounded, and this often includes a model-based sensitivity analysis to adjust for the imbalance.  In the case of Makena, a drug receiving accelerated-approval to reduce pre-term births, a predictive covariate was imbalanced in a moderately-large (n=453) randomized trial\cite{meis2003prevention}.  A post-approval confirmatory trial did not confirm Makena’s efficacy\cite{chang2020withdrawing}, and the imbalanced covariate was described as ‘problematic’ when the Food and Drug Administration deliberated whether to recommend Makena be withdrawn from market\cite{makena2022review}.  In another case, a trial to test the survival benefit of salmeterol plus fluticasone propionate among patients with chronic obstructive pulmonary disease ended just shy of statistical significance (p = 0.052)\cite{Calverley:2007gx}.  Equipoise on the intervention's benefit remained with a lament that "given the disappointment and expense of conducting the TORCH study, it is unsurprising that funders have been reluctant to revisit this topic."\cite{Calverley2021} Plausibly, the trial may have been more conclusive with a more efficient study design.

The past two decades have seen major advancements in CAR schemes and inferential theory. The use of a distance matrix has been incorporated into Matched Randomization (MR)\cite{Greevy:2004ke, GreevyJr:2012hp} and Sequential Matched Randomization (SMR) \cite{Kapelner:2014cu, Kapelner2021}, Rerandomization\cite{Morgan:2012iq,Zhou2018}, and Pairwise Sequential Randomization (PSR) \cite{Qin2016}. New theory guides parametric-based hypothesis testing following covariate-adaptive randomization -- schemes which dynamically randomize sequential enrollment to minimize a balance metric\cite{Shao2010a,Shao:2013jz, Ma2020}.  Yet, despite these advances, traditional stratified randomization remains more frequently used than alternative CAR schemes \cite{sverdlov2023randomization, mcpherson2012use}.  Traditional stratified randomization is an easy-to-implement method yet is well understood to be limited by its ability to adjust for only relatively few, categorized covariates\cite{therneau1993many}.

A form of stratification, MR randomizes within an optimal set of two-participant matches adjusting for any number of continuous or categorical covariates.  The set of optimal matches are those which result in the smallest sum of paired distances calculated from a distance matrix.  In trials with sequential enrollment, SMR uses the "matching-on-the-fly" algorithm to match participants with the first unmatched participant that meets a pre-specified similarity threshold. Because participants are matched in pairs, MR and SMR may be considered personalized randomization schemes and is our primary focus in this work. Optimizing matched pairings distinguishes MR/SMR from Rerandomization and PSR which also use a distance matrix though to minimize between-arm covariate imbalance. Open questions for SMR are how to relax the requirement of pre-specifying the similarity threshold and recover some of the optimality in pairing matches under sequential enrollment.

At the point of inference, it is considered desirable for a randomized trial to provide a marginal, unadjusted analysis to test treatment efficacy\cite{Freedman:2008eq,Lin:2013jh}.  The common, likelihood-based population model (e.g., two-sample t-test) assumes participants are drawn as a simple random sample from an infinite or sufficiently-large population\cite{Abadie2020a}.  Uncertainty comes from the sampled participants.  In contrast, the Neyman Model assumes a finite population and that uncertainty comes from the potential outcomes observed due to treatment assignment\cite{rosenberger2015randomization}.  Under this model, and assuming a constant null treatment effect, Randomization-Based Inference (RBI) provides an exact test of the treatment effect by comparing the extremity of an observed statistic to all possible observations of the statistic from each randomization sequence. The RBI null distribution is asymptotically normal when randomizing with Complete Randomization (CR) in which each participant has an equal and unrestricted probability of treatment assignment. Despite this connection between parametric and randomization-based inference, an argument is made that the Neyman model is most appropriate for a randomized trial with enrollment constraints\cite{Rosenberger2019}.  Yet, another appeal of the Neyman model is in using RBI as an exact test of the marginal treatment effect. It is free of modelling assumptions and efficiency gains may be observed following CAR as compared to following CR. The Food and Drug Administration has relied upon RBI in evaluating therapeutic efficacy \cite{U.S.FoodandDrugAdministration2017}. (See also the trial GOG-0218. Initial trial results did not use RBI\cite{burger2012}. However, RBI was the primary analysis for regulatory evaluation as highlighted in Table 17 of Avastin's prescribing information\cite{avastin2018}). 

The objective of this work is to extend and evaluate the performance of SMR schemes.  Proposed extensions include: (i) allowing batch enrollment with simultaneous matching and randomization, (ii) using a dynamic matching threshold with control for chronological bias, and (iii) allowing matches to break and rematch (Sequential Rematched Randomization; SRR).  We aim to address the questions: Do the SMR extensions improve the covariate balance and efficiency compared to SMR without extensions?  Do MR and SMR schemes share a common limitation of stratified randomization -- that adjusting for more covariates can be detrimental to covariate balance and/or efficiency?  We will use simplified simulations (under the population sampling model) and data from the REACH trial as a case-study (under the Neyman model) to address these questions.

As a secondary objective, we will use the case study to investigate two additional questions.  First, as a side-by-side frame of reference for this particular study, how do SMR schemes compare in covariate balance and efficiency to other CAR schemes (stratified, PSR, and Atkinson's minimization\cite{Atkinson:1982kt})?  Second, can adjusting for covariates in randomization (i.e., CAR + RBI) be as powerful as adjusting for covariates in a regression model (i.e., CR + Covariate Adjusted Regression Model)? The second question is motivated from a common analysis strategy of not adjusting for covariates in randomization, reporting a primary analysis of a parametric test of the marginal treatment effect, and performing a secondary analysis which adjusts for covariates in a parametric model.

The work continues as follows.  First, we will introduce notation used throughout and describe the matching-on-the-fly algorithm.  Second, we will propose the SMR extensions.  Third, we will compare SMR with and without extensions in simplified settings that vary sample size,  covariate distribution, and covariate association with outcome.  Fourth, we will present the case study to compare SMR schemes and address the secondary questions of interest. Sections three and four will end with conclusions which will be revisited in a final discussion section.

\section{Notation and Matching-on-the-fly algorithm}

A study enrolls $i \in \left\{1,\ldots,N\right\}$ participants throughout $b \in \left\{1,\ldots,B\right\}$ batches of sequential enrollment. At any point in time the total number of enrolled patients is $n \in \left\{1,\ldots,N\right\}$. While the planning for $N$ is fixed, $N$ may be random due to enrollment logistics. Each participant is assigned to $W_i \in \left\{0, 1\right\}$, respectively denoting control or treatment assignment, and has potential outcomes of $Y_i(W_i)$.  A "sharp" treatment effect is defined as $Y_i(1) = Y_i(0) + constant$, which under the null hypothesis implies $Y_i(1) = Y_i(0)$. The observed randomization sequence is $\boldsymbol{W}_N \coloneqq \left\{ w_1, \dots, w_N \right\}$, and $h(\boldsymbol{W})$ is an observed statistic such as the difference in treatment means. Let $\Omega_{RS}$ denote the set of randomization sequences under a given randomization scheme and $||\Omega_{RS}||$ be the cardinality.  A two-sided RBI p-value is defined as $\frac{1}{||\Omega_{RS}||} \sum \limits_{\boldsymbol{W}^{\ast} \in \Omega_{RS}} | h(\boldsymbol{W}^{\ast}) | \ge | h(\boldsymbol{W}) |$. When observing the full set $\Omega_{RS}$ is computationally intractable, it may be approximated through equiprobable monte-carlo sampling from $\Omega_{RS}$\cite{Rosenberger2019}. A CAR scheme adjusts for $p$ covariates which may include dummy variables for categorical covariates.

SMR uses the matching on-the-fly algorithm to form two-person pairs (a.k.a. matches/strata) of participants and to randomize treatment within pairs\cite{Kapelner:2014cu}. An initial set of participants are randomized using CR (i.e., drawn independently from $Bern(0.5)$) and form a “reservoir” of randomized but unmatched participants. In fully sequential manner (i.e., $B=N$), the pairwise similarity between an enrolling participant and reservoir members is calculated as a distance obtained from a distance matrix. A match forms when the distance between the enrolling participant and their best match is less than a pre-specified matching threshold -- a quantile of randomly matched distances. This reference distribution is denoted $F$ and $F_b$ following a given batch of enrollments. The entering participant receives the opposite treatment to their match. Otherwise, the entering participant is randomized using CR and enters the reservoir. By the end of the study, some participants may not have matched, and treatment allocation may be imbalanced. Conditioning on matches, each participant has an equal probability of being randomized to each arm.  The SMR algorithm has been extended to prioritize covariate balance based upon their association with the outcome\cite{Kapelner2021}, though conditioning on outcomes is outside the scope of this work.

Common SMR practice is to use Mahalanobis Distance as the distance matrix and to build the reservoir up to $p+2$ to allow for the earliest possible matching.  $F_b$ is commonly estimated as $\hat{F_b}$ using the parametric $F_{(p, n-p)}$ distribution, though $\hat{F}_b$ may be estimated empirically by bootstrap sampling a random set of $n/2$ matches from the upper- (or lower-) triangle of the distance matrix.  We use the notation SMR(Q, E) and SMR(Q, F) respectively to denote using SMR with matching threshold set as the quantile Q of $\hat{F}_b$ estimated either empirically or by $F_{(p,n-p)}$.

For the proposed extensions, let $U_b$ denote the set of unmatched participants including the enrolling participant and $||U_b||$ the cardinality.  Let $||R_b||$ be the number of expected remaining study entrants.

\section{Extensions to SMR}

\subsection{Batch matching}
SMR requires participants to be randomized one at a time. A fully sequential matching scheme "greedily" finds the best matches at the moment although a better match may later exist.  In contrast, MR randomizes all participants in a single batch (i.e., non-sequential enrollment) within the set of matches that minimize the total sum of matched distances.  Applying this optimality criteria, we extend SMR for randomization of any batch size.  As multiple participants enroll, potential two-person matches are found which minimize the total sum of matched distances.  Randomization occurs within matches that meet the matching threshold.  
MR is a special case of SMR with $B=1$ and is the least greedy of SMR schemes.

\subsection{Dynamic matching threshold with chronological bias control}
\label{subsec:threshold}

Determining an appropriate fixed matching threshold may be challenging. An overly strict fixed matching threshold would yield no matches and in turn may be no better than CR. In contrast, an overly relaxed matching threshold would degenerate to a block-two randomization scheme vulnerable to subversion bias and no longer adjust for covariates. We propose a dynamic matching threshold which reflects the chance of matching to an existing reservoir member versus a future enrolling participant. Formally, this is the quantile $Q_b$ of $\hat{F}_b$ where
\begin{equation*}
Q_b=\frac{||U_b||-1}{||U_b||+||R_b||-1}.
\end{equation*}
When estimating $\hat{F}_b$ with bootstrap sampling, the dynamic matching threshold is the averaged $Q_b$ percentile across bootstrap samples of $\hat{F}_b$. To achieve equal treatment allocation, the matching threshold may be completely relaxed, by setting to infinity, when the reservoir size is the same or less than the number of participants left to enroll.  The dynamic matching threshold is then:
\begin{equation*}
Threshold_b=\left\{\begin{matrix}{\hat{F}}_b^{-1}(Q_b)&||U_b||<||R_b||\\ \infty &||U_b||\geq||R_b||\\\end{matrix}\right..
\end{equation*}
To control for potential chronological bias, we implement SMR using a Maximum Tolerable Imbalance (MTI) procedure, which is a class of procedures that ensures the allocation imbalance is no worse than a pre-specified imbalance \cite{Zhao2014}. Here, we've used the Big Stick Design which forces assignment treatment allocation as needed to maintain the allocation within the tolerable imbalance\cite{soares1983some, berger2021roadmap}. When a batch of partcipant(s) enroll, the first step is to identify any matches and make appropriate treatment assignments. The remaining participants are then randomized using the first set of random CR assignments in which the overall and batch allocation imbalance is within the MTI limits. In some instances, it may not be possible to achieve both the overall and batch imbalance. Priority is given to achieving overall allocation within the MTI while then getting as close to the batch MTI as possible.

\subsection{Sequential Rematching}
Throughout enrollment, SRR allows matches to break and rematch to participants who do not share the same treatment assignment.  It aims to recover some of the optimality of MR matches.  We call the SRR algorithm rematching on-the-fly or sequential rematching with the allowance of batching. Operationally, the distance matrix is calculated for all participants -- including those on study and to be randomized. The objective remains to minimize the sum of matched distances among possible matches. To prevent matching within the same treatment arm, the corresponding paired distances are set to $\infty$ (or an impractically large number). If the new participant(s) are matched to an existing participant, they receive the opposite treatment arm. Otherwise, they receive a random treatment assignment. SRR could be carried out with a pre-specified fixed treatment threshold with CR or with the dynamic threshold having control for chronological bias as described in Section~\ref{subsec:threshold}.

\subsection{Practical implementation: nbpMatching and missing baseline characteristics}

The nbpMatching package\cite{beck2016nbpmatching} can be used to operationalize SMR and SRR with batch enrollment. The function \texttt{nbpMatching::gendistance} calculates and prepares the distance matrix under SMR or SRR constraints, and \texttt{nbpMatching::nonbimatch} subsequently finds optimal matched pairs meeting a specified threshold. SMR constrains possible matches by requiring previous matches to remain intact through the \texttt{force} input to \texttt{nbpMatching::gendistance}. This sets the distance of existing matches to 0. SRR constrains possible matches by excluding matches within the same treatment arm using the \texttt{prevent} input of \texttt{nbpMatching::gendistance}. This sets the distances of same-arm participants to $\infty$. SRR keeps \texttt{force} at the default -- NULL -- to consider all other possible matches. Calling \texttt{nbpMatching::gendistance} without constraints and \texttt{nbpMatching::nonbimatch} without a threshold results in MR. All code used for this paper is available in the supplement. The code for the real-world data application in Section~\ref{sec:reach}, is generalized for use with other trial data.

As common with CAR schemes, SMR and SRR require a complete set of baseline covariates to carry out randomization; yet, this may not always be feasible.  For example, a predictive baseline lab value may take more time to process than desired for randomization. While there are many viable options for imputation, the nbpMatching R package may also be useful for MR and SMR schemes, here by using predictive mean matching from other covariates.  We do not attempt to investigate the performance of imputing missing baseline covariates in this work.  However, we mention the feature of imputing missing baseline covariates because it played a critical role when SMR was used in the real-life application presented later.

\section{Simulations under simplified settings}

\subsection{Simulation set-up}

In simplified settings that condition SMR on two covariates, we sought to find fixed matching thresholds that yielded optimal covariate balance and estimation efficiency and to assess the benefit of the proposed extensions.  We hypothesized that (H1) the optimal, SMR fixed matching threshold would be sensitive to covariate distribution, covariate association with outcome, and sample size and that (H2) each extension, individually and potentially collectively, would improve covariate balance and estimator efficiency.

Under the population sampling model, standard normal outcomes were simulated such that two covariates each explained 10\% (mild) and/or 25\% (moderate) outcome variability.  The first of the covariates was standard normal, and the second followed one of three distributions (CD1) dichotomous with 20\% prevalence, (CD2) dichotomous with 50\% prevalence, and (CD3) standard normal.  For each covariate distribution setting, we considered four covariate predictive settings for the two covariates: (CP1) moderate and moderate, (CP2) mild and moderate, (CP3) moderate and mild, (CP4) mild and mild.  Setting CD1+CP3 reflects situations when there is a moderately-associated continuous covariate and a mildly-associated, and somewhat rare, dichotomous covariate.  

Each combination of settings were evaluated with $N$ of 50 and 250 for a total of 24 simulation settings.  Each setting was replicated to generate 150K datasets.  CR, MR (the upper limit potential of batched SMR), SMR and SRR were carried out for each simulated dataset.  Fixed matching thresholds ranged from 5\% to 50\% of the $F(p, n-p)$ distribution.  The dynamic threshold included an MTI of 4.  Balance was measured as two metrics. The first, the sum of matched distances, is the metric optimized by MR. Any unmatched participants following SMR were randomly matched to contribute to the sum of distances. Second, balance was measured as the average absolute standardized mean difference (SMD) of each covariate across replicates.  While the former gives insight into the quality/refinement of matches, the later is a between-arm measure of covariate imbalance similar to that observable from a table of baseline demographics. Estimator efficiency is measured as the standard deviation of the difference of outcome means estimated from each replicate.

\subsection{Results}

\begin{figure}[H]
\centering
\includegraphics[width=1\columnwidth]{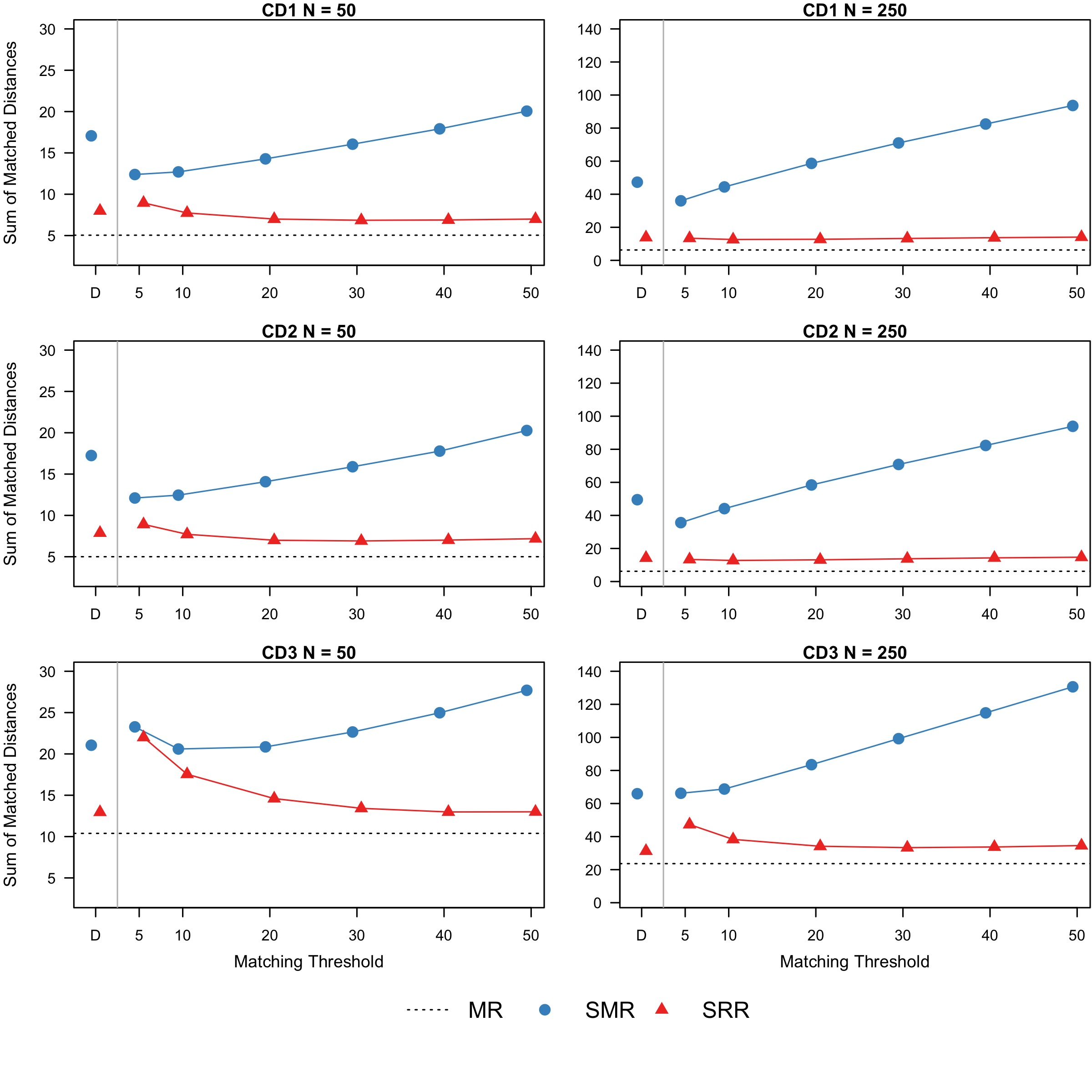}
\caption{Match quality: The sum of matched distances averaged across replicates when using SMR with and without extensions under simplified simulation settings. 'D' on the x-axis denotes performance under the Dynamic threshold. MR minimizes the sum of matched distances and provides a bound to this metric.  Some optimality is lost when participants enter sequentially.}
\label{fig:sim_sumdistances}
\end{figure}

\begin{figure}[H]
\centering
\includegraphics[width=1\columnwidth]{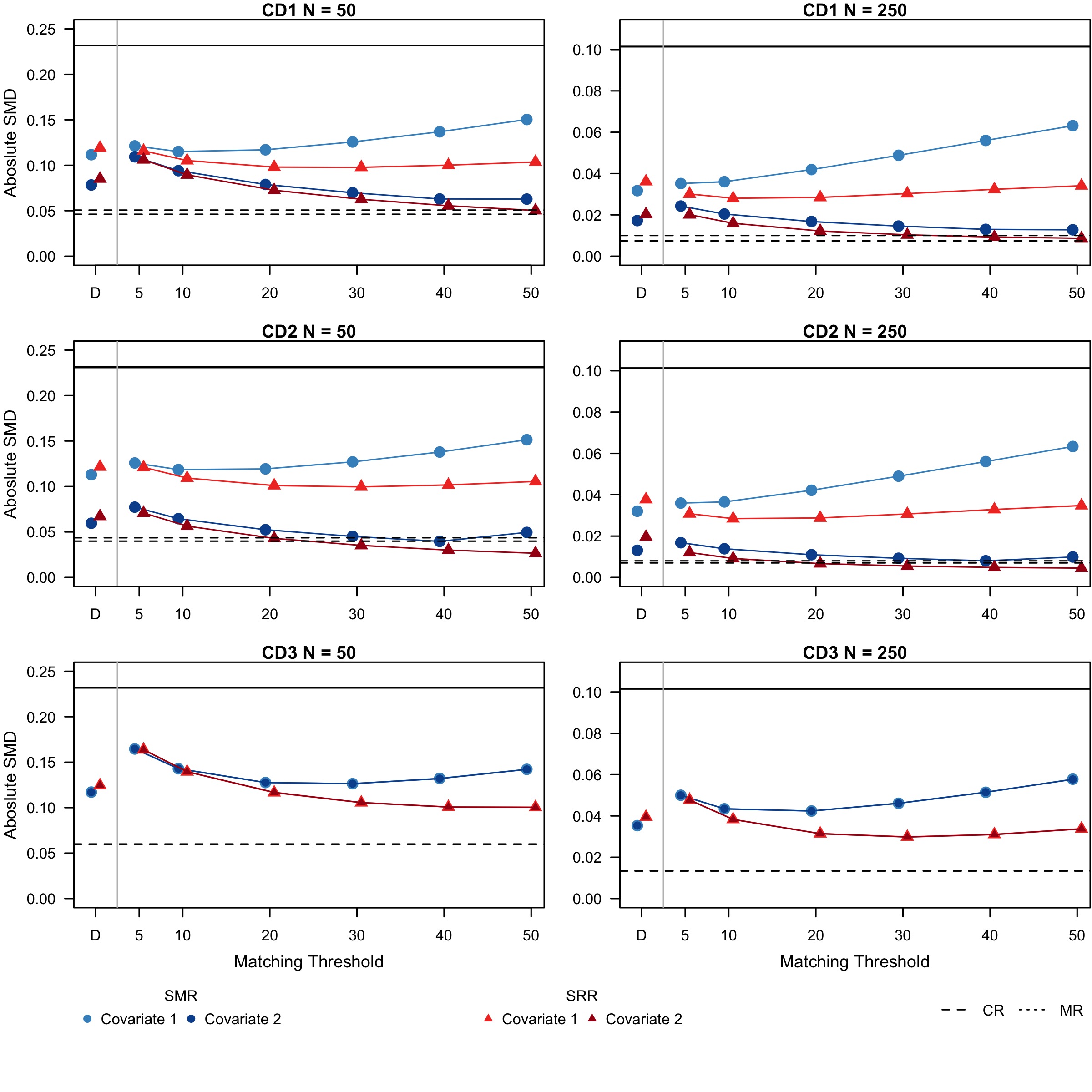}
\caption{Between-arm covariate balance: The average absolute Standardized Mean Difference under simplified simulation settings.  Covariates are distributed as: CD1 -- N(0, 1) and Bin(.2); CD2 -- N(0, 1) and Bin(.5); and CD3 -- N(0, 1) and N(0, 1).  In both CD3 figures, Covariate 1 has the same distribution as Covariate 2 and is plotted behind Covariate 2.}
\label{fig:sim_balance}
\end{figure}

\begin{figure}[H]
\centering
\includegraphics[width=1\columnwidth]{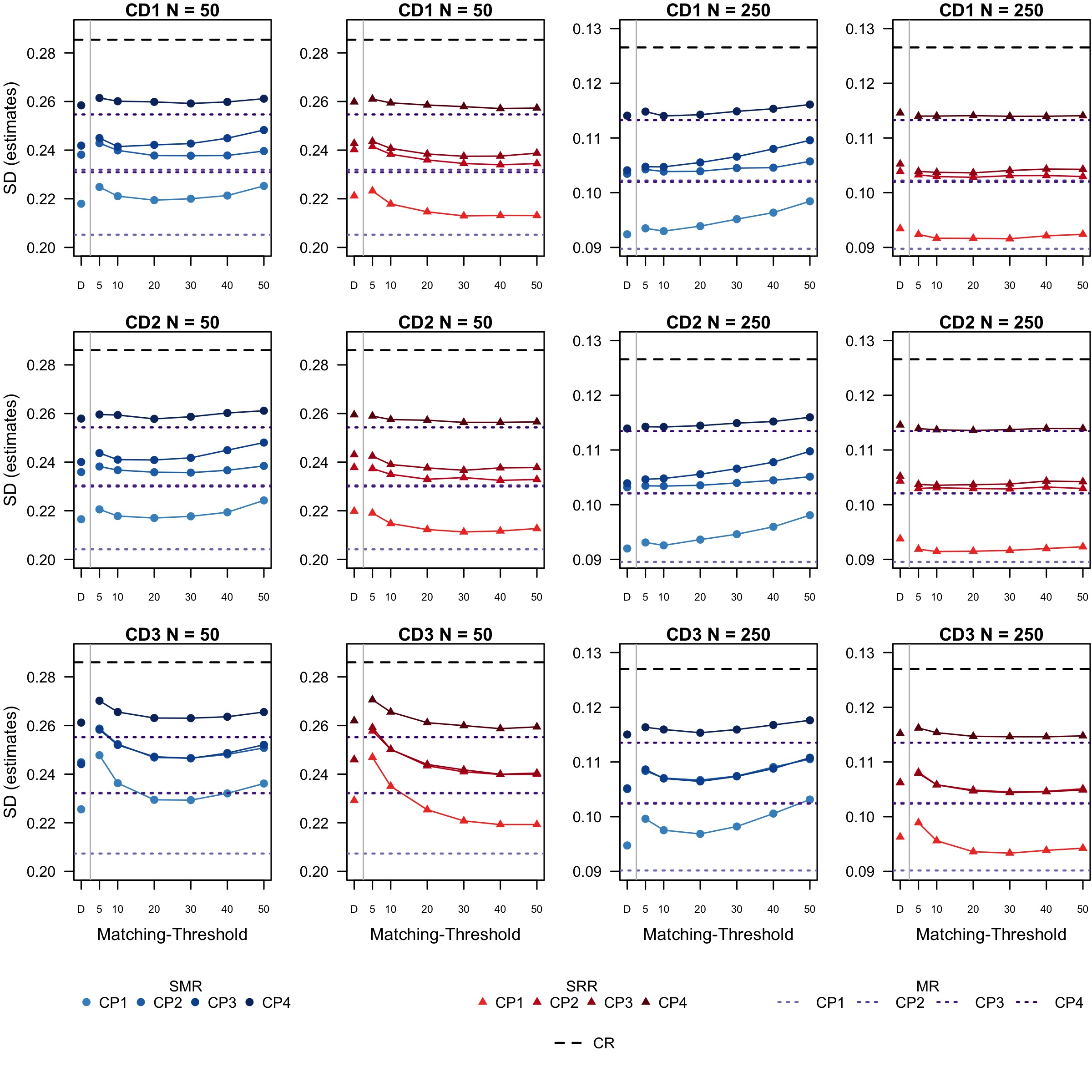}
\caption{Estimator efficiency: The standard deviation of estimates under simplified simulation settings.  The predictiveness of each of the two covariates upon outcomes, in terms of R$^2$ are denoted as: CP1 -- 25\% each; CP2 -- 10\% and 25\%; CP3 -- 25\% and 10\%; and CP4 -- 10\% each. Covariates are distributed as: CD1 -- N(0, 1) and Bin(.2); CD2 -- N(0, 1) and Bin(.5); and CD3 -- N(0, 1) and N(0, 1).}
\label{fig:sim_eff}
\end{figure}

Covariate-balance and estimator efficiency results are presented in Figures ~\ref{fig:sim_sumdistances} (sum of matched distances), ~\ref{fig:sim_balance} (absolute standardized mean difference), and ~\ref{fig:sim_eff} (estimator efficiency).

H1 (SMR with fixed thresholds): Under SMR, there were U-shaped balance and efficiency curves (CD3, N=50) that became less sensitive with increased sample size and/or in the presence of categorical covariates. 

Under SMR, the best fixed threshold for minimizing the sum of distances was the smallest threshold (with exception for CD3 N=50 in which a 10\% matching threshold was optimal).  For between-arm covariate balance, the optimal matching threshold for continuous covariates (CD3) decreased with sample size from about 30\% (N = 50) to about 20\% (N = 250) to about 10\% in the presence of a categorical covariate (CD2, N =  250).  The optimal matching threshold for balancing categorical covariates generally remained at 30-50\% (CD1 and CD2).  For estimator efficiency, it was worse to have an extreme large matching threshold than an extreme small matching threshold (i.e., 50\% vs 5\%; exception CD3 with n = 50).  Having a categorical covariate seemed to taper the diminished returns of a large matching-threshold (for example 50\% matching threshold of CD1 and CD2 vs CD3).  

H2 (SMR with dynamic threshold): When both covariates were continuous and for all metrics, SMR with a dynamic threshold was as good as any fixed threshold regardless of sample size. In the presence of a categorical covariate, a fixed threshold of 10\% and smaller was better than a dynamic threshold regardless of sample size for the two balance metrics.  SMR with a dynamic threshold was more efficient across all covariate distributions and predictiveness patterns. 

For between-arm covariate balance of a continuous covariate (Covariate 1) SMR performed better with a dynamic matching-threshold than with fixed matching thresholds.  For between-arm covariate balance of a categorical covariate, SMR with a dynamic matching-threshold was at least comparable and more often superior to using a fixed matching-threshold (which was best in the range of 30\%-50\%).

H2 (SMR vs MR and SRR): MR was superior to SRR then SMR in minimizing the sum of matched distances. Between-arm covariate balance and efficiency was best under MR, followed by SRR, then SMR for a given covariate distribution, covariate association, sample size, and fixed matching-threshold.  When using a dynamic threshold the ranking changed to MR, SMR, then SRR.  Also, when balancing between-arm categorical covariates in CD2, SRR surpassed MR when using a fixed matching threshold of at least 30\%.

\subsection{Conclusions}

In general conclusions to these simplified settings, the optimal, fixed matching threshold for SMR was sensitive to covariate distribution and association upon outcome, especially for smaller sample sizes and in the absence of categorical covariates.  In these simulations, a fixed matching threshold of about 10-20\% was generally a sweet spot for optimizing SMR between-arm covariate balance and estimator efficiency.  SMR with a dynamic threshold achieved competitive sum of matched distances, between-arm covariate balance, and efficiency as the optimal SMR fixed thresholds. The optimal performance under SRR tended to be better than the optimal performance under SMR; and batched randomization tended to remain superior to all.  These conclusions were drawn under simplified settings and the recommendation may change in practical situations with different covariates and sample sizes.

Based on these results, it may be tempting to consider SRR with a high, fixed threshold.  However, the U-shaped efficiency curve for SRR (e.g., CD3 with N = 250) suggests this may be overall counterproductive. And, a fully relaxed matching threshold would degenerate to a deterministic minimization algorithm.  The dynamic threshold protects against this risk.

\section{Application to REACH trial}
\label{sec:reach}

\subsection{Background}

The Rapid Education/Encouragement And Communications for Health (REACH)  randomized clinical trial provided text message-delivered diabetes support to help adults with type 2 diabetes improve glycemic control and adhere to treatment medication. 512 Participants enrolled and participants were randomized across three treatment arms with a 2:1:1 allocation ratio: enhanced treatment as usual (a text message when study A1c results are available), frequent diabetes self-care support text messages, and frequent diabetes self-care support text messages with monthly phone coaching.  At baseline, clinical and demographic covariates were collected that may be associated with the outcome 12-month Hemoglobin A1c (HbA1c), including site (community clinic versus Vanderbilt University Medical Center primary care), HbA1c, age at enrollment, gender, years of education, years since diabetes diagnosis, race / ethnicity, medication type, income, and type of health insurance. A difference in HbA1c favoring the REACH arm was found at 6 months\cite{Nelson2021}.

Randomization often occurred before all baseline covariates could be collected. Notably, baseline HbA1c generally took additional processing time. In this study, clinical coordinators enrolled participants and sent baseline data to the statistician for weekly batch randomization via REDCap \cite{harris2009metadata,harris2019redcap}.  Using the nbpMatching package, missing baseline covariates were imputed and SMR with batch randomization was utilized.  Once HbA1c was available, it was updated and included in future SMR randomizations.

With a complete set of REACH baseline characteristics, we wished to address (i) how SMR with and without extensions could improve the covariate balance and study efficiency and (ii) whether adjusting for additional covariates impacts performance. Balance would be measured by the observed matched distances and average absolute standardized mean differences.  Study efficiency would be measured as the power to reject a sharp treatment effect upon 12 month HbA1c of -0.5 (i.e., beneficial), and estimator efficiency would be measured as the variance of the estimates.  

Secondarily, we wish to (i) provide a frame of reference of how SMR schemes compare to other CAR schemes and (ii) see whether CAR with RBI could be as powerful and efficient as CR (i.e., no randomization covariate adjustment) with model-based covariate adjustment.  Other CAR schemes were chosen for the reasons: Atkinson's D$_A$ Biased Coin Design (D$_A$-BCD) is a canonical approach to covariate-adaptive randomization; PSR is a covariate-adaptive randomization which uses Mahalanobis Distance; and stratification is a common strategy for CAR.


\subsection{Simulation set-up}

For simplicity, we supposed the study was an equally allocated trial comparing usual care to either of the intervention arms.  The base SMR comparators included SMR(20, E) and SMR(20, F).  We carried out SMR extensions as SMR(D, E), SMR(D, F), SRR(20, E), SRR(20, F), SRR(D, E), and SRR(D, F). MR was included to reflect the potential of sequential batched randomization and the potential loss under fully-sequential matching. Priority covariates included baseline HbA1c, medication type, and time since diabetes diagnosis (R$^2$=0.26 with 12-month HbA1c) while all covariates were those aforementioned (R$^2$=0.32 with 12-month HbA1c). We used 200 bootstrap samples for estimating $\hat{F}_b$ empirically.

 The D$_A$-BCD regression model was pre-specified to include first-order covariate adjustments, and used block two randomization until there were more participants than degrees of freedom. (See supplement for consideration of using the generalized inverse\cite{Senn:2010bg}).  We used the PSR coding made available in existing literature\cite{Ma2020}.  For D$_A$-BCD and PSR, we used a biased coin of 0.75.  Stratified randomization used clinically-meaningful categorized derivations of baseline HbA1c (<7.0\%, 7.0-8.0\%, 8.0\%+), age ($\le$ 60, >60), years of education ($\le$12, >12), and time since diabetes diagnosis (<10, $\ge$10 years). 

For each randomization scheme, we kept baseline covariates and enrollment fixed as observed in the REACH trial and generated random outcomes under the potential outcomes framework (i.e., the Neyman model).  First, using monte-carlo draws from each randomization scheme, we generated 500K randomization sequences.  Then, for each randomization sequence, we used observed 12-month HbA1c as the outcome of interest and imposed a sharp treatment effect of 0 (null hypothesis) and -0.5 for participants randomized to treatment -- yielding 500K observed potential outcome datasets.  For each outcome dataset, we recorded the difference in average treatment effect.  We tested the average treatment effect using RBI, a two-sample t-test, and ordinary least-squares regression model adjusting for covariates.  For computational feasibility, we chunked the simulation set-up into batches of 25K monte-carlo replicates each.

Between-arm covariate balance was evaluated as the average absolute SMD of all covariates.  With alpha of 0.05 and a two-sided test, Type I error and power were estimated as the proportion of outcome datasets that rejected a 12-month treatment difference in HbA1c for under each inferential test. Relative study efficiency was estimated as the effective gain in sample relative to power achieved under CR with t-test inference. Estimator efficiency was estimated as the variance of the estimated mean difference in 12-month HbA1c across the outcome datasets, and efficiency was again compared relative to CR with t-test inference. For SMR schemes, we further assessed study- and match-level balance through the resulting matched distances. Any unmatched participants resulting from a fixed threshold were randomly matched to obtain a distance. We averaged the observed ranked distances across replicates. For a frame of reference, the average ranked distance was compared to the empirical distribution of randomly matched distances. We called this "match quality."

Acknowledging that in practice most designs restrict for chronological bias using a blocking strategy, we repeated the simulation using the Big Stick Design MTI design. We also considered non-linear adjustments to the regression model. Results of these schemes are reported in the supplement.



\subsection{Results}

SMR(E, 20) achieved greater between-arm covariate balance, power, and estimator efficiency than SMR(F, 20).  For simplicity, we focus on SMR results which estimated $\hat{F}_b$ empirically. Empirical estimation of $\hat{F}_b$ provided more consistent performance for a given scheme and, in most cases, was superior to estimation using $\hat{F}_b$ as $F_{(p,n-p)}$ (see supplement for full results). 

\begin{figure}[H]
\centering
\includegraphics[width=1\columnwidth]{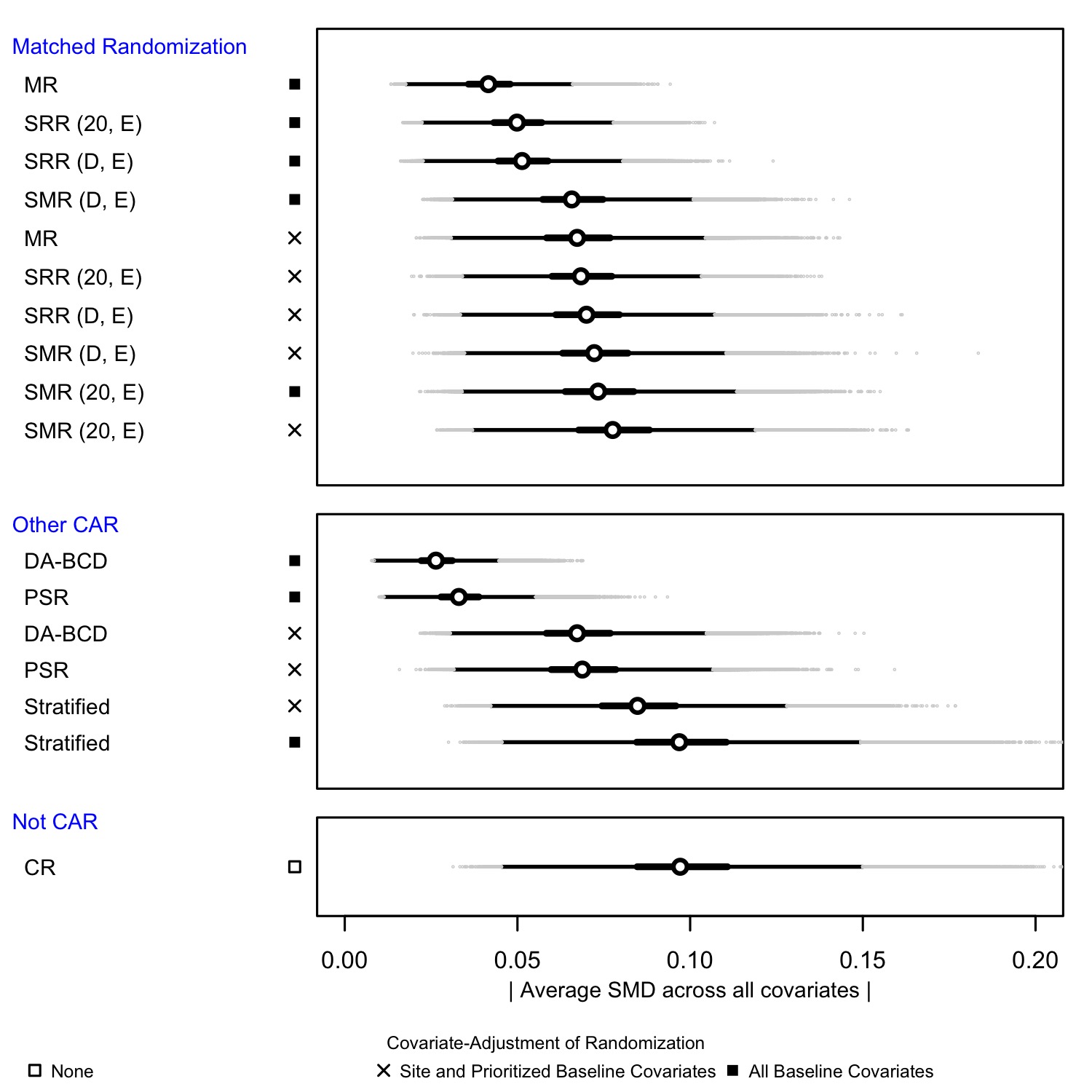}
\caption{Between-arm Covariate Balance: Boxplot of average absolute Standardized Mean Difference (SMD) across all baseline covariates of interest in REACH trial after Matched Randomization, Other CAR, and Not CAR schemes.  A lower SMD is better.}
\label{fig:bal}
\end{figure}

Between-arm covariate balance (Figure \ref{fig:bal}): Conditioned on all covariates, the average balance across all covariates was 0.07 under SMR(20, E).  Each extension, individually and collectively, improved upon this measure of covariate balance with MR and SRR schemes achieving greater balance than SMR schemes. Unlike traditional stratified randomization, SMR schemes achieved greater overall covariate balance when conditioning on all versus priority covariates.

\begin{figure}[H]
\centering
\includegraphics[width=1\columnwidth]{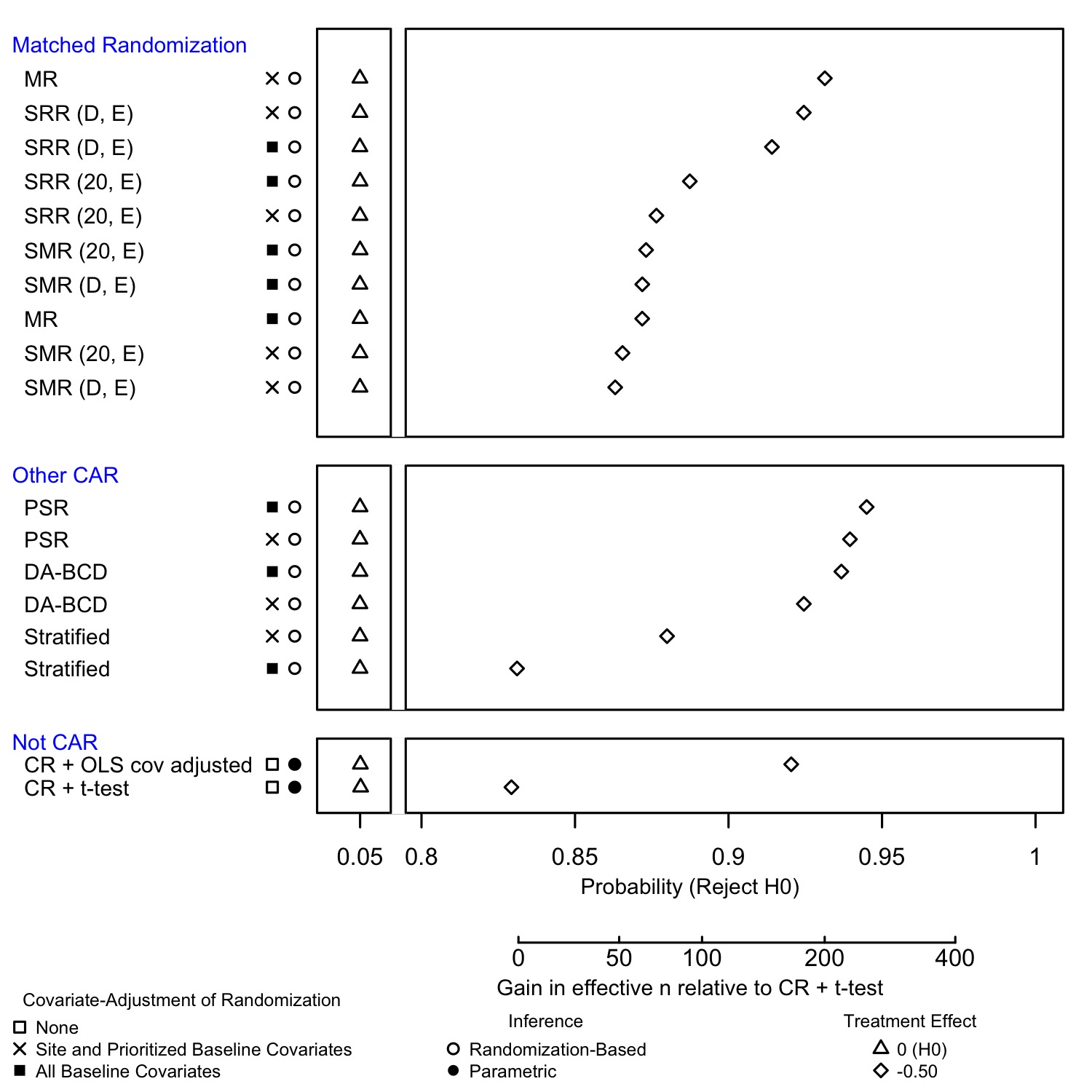}
\caption{Type I Error and Power to test the Average Treatment Effect: The estimated probability of rejecting the null hypothesis under a sharp treatment effect of 0 (null) and -0.5 (highly meaningful) in the REACH trial after Matched Randomization, Other CAR, and Not CAR randomization schemes.  Power is calculated under RBI except for Not CAR schemes which use the population model for inference.  Under the sharp treatment effect of -0.5, gains in effective sample size are relative to CR.  For example, this case study was carried out with 512 enrolled participants yet when designed with SRR (D, E) and analyzed using randomization based inference, the power is as though an additional 177 participants were on study under CR and when analyzed using an unadjusted population model.}
\label{fig:pow}
\end{figure}

Power and estimator efficiency (Figure \ref{fig:pow}): SMR without extensions rejected the null hypothesis under RBI 87\% of the time when conditioning on priority and on all covariates.  All extensions improved upon SMR(20, E) conditioned on priority covariates. One exception was SMR(D, E) conditioned on priority covariates which was equivalent to SMR(20, E) within rounding to the nearest hundredth. Conditioning on all covariates improved the power of SMR(20, E), SMR(D, E), and SRR(20, E); however, SRR(D,E) and MR achieved greater power when conditioning on priority covariates. Though not always the case, the more powerful schemes also tended to have more efficient estimators (supplement).  Differences in rank ordering, and in relative performance, might be attributable to differences in scheme's RBI reference distribution (for example, SRR schemes had greater kurtosis than SMR schemes).

Side-by-side comparisons: As a frame of reference, PSR and D$_A$-BCD schemes achieved as good, and in some cases, superior covariate balance and power than the best performing SMR scheme -- SRR(D,E) on priority covariates. In these two other CAR schemes, conditioning on more covariates further increased between-arm balance and power. As expected, stratifying on all covariates essentially mirrored CR performance. Stratifying on priority covariates still substantively improved between-arm covariate balance, power, and estimator efficiency.

In this case study, testing the marginal, average treatment effect was more powerful using a CAR scheme with RBI than using CR with a t-test. Certain CAR schemes with RBI were also more powerful than CR with a regression model adjusting for all covariates.  These schemes included SRR(D,E) on priority covariates, PSR, and D$_A$-BCD.  The resulting power of these designs were as though at least 177 extra participants were on study supposing CR was used for randomization and the t-test was used for analysis. These general conclusions held for using a randomization scheme to account for chronological bias and when flexibly modelling continuous covariates to have non-linear associations upon outcome (supplement).

\begin{figure}[H]
\centering
\includegraphics[width=1\columnwidth]{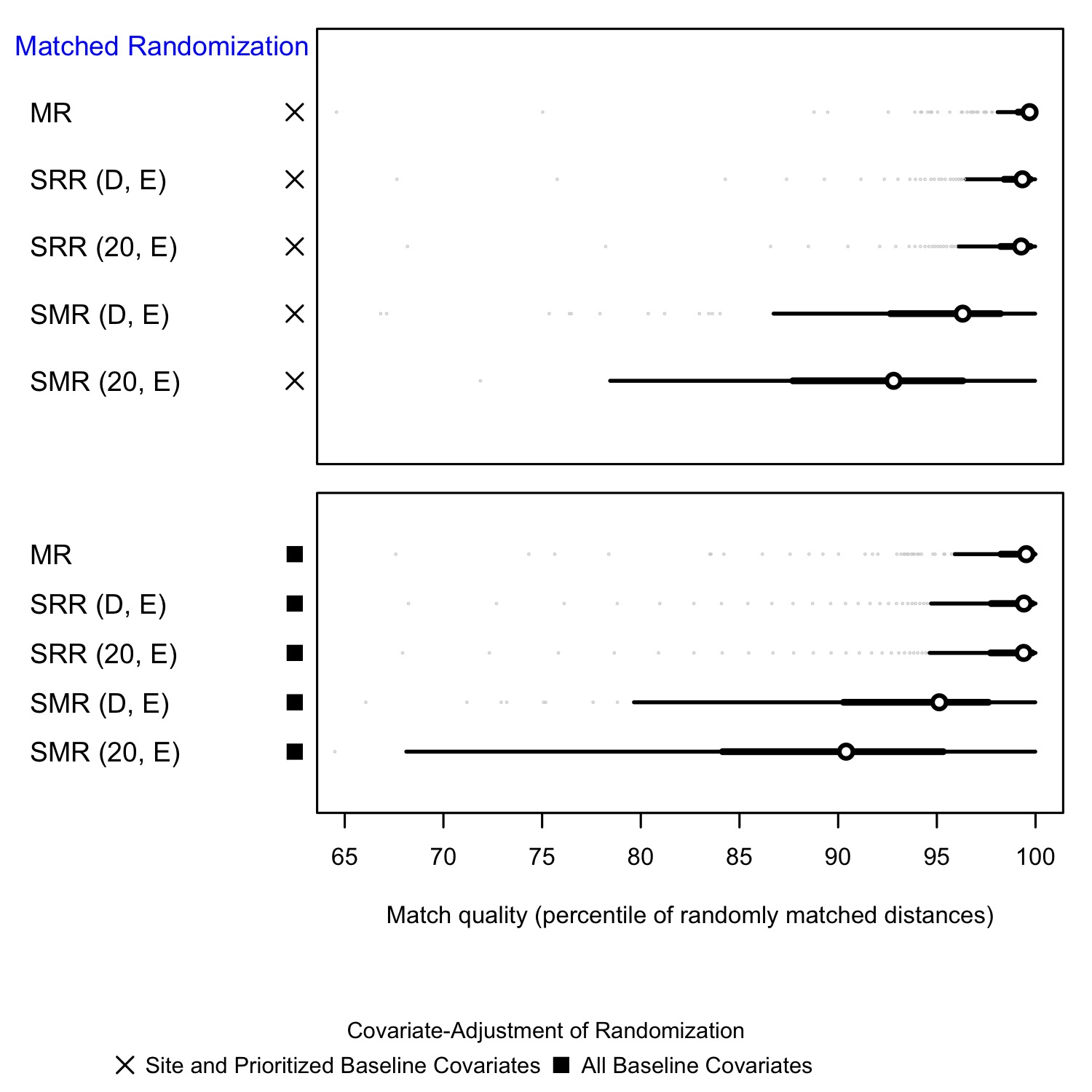}
\caption{Match Quality: Distribution, presented as a boxplot, of match distances following a matched randomization scheme in the REACH trial. Distances are presented as a percentile of randomly paired distances (i.e., $\widehat{F}$ estimated empirically). Unmatched participants in SMR(20, E) and SRR(20, E) are randomly matched and distances recorded.}
\label{fig:dists}
\end{figure}

Match quality: Whether conditioned on priority or all covariates, each SMR extension improved the quality of matched pairs (Figure \ref{fig:dists}).  SRR schemes nearly recovered the full quality of matches optimally found under a single-batch MR. When conditioned on priority covariates, 98\% of SRR(D, E) matched pairs distances were better than the top 10 percentile of randomly matched distances.

\subsection{Conclusions}

While exact results may differ from study to study, we observed SRR with a dynamic threshold to improve the quality of matches, between-arm covariate balance, power, and estimator efficiency over SMR with a fixed threshold of 20\%. In contrast to stratification, SMR schemes achieved greater between-arm covariate balance when adjusting for all covariates than when adjusting for priority covariates. The impact of adjusting for additional covariates upon power was not universal across SMR schemes. Adjusting for more covariates was favorable for SMR schemes and SRR(20, E), though unfavorable for SRR(D,E) and MR.

Among the side-by-side CAR comparisons, each class of schemes was optimal in different metrics. D$_A$-BCD achieved the best between-arm covariate balance, PSR achieved the best RBI power, and only matching schemes could provide insight into how personalized randomization was to the participant. Depending on a trial's objective, each of these metrics may have particular appeal.

The REACH trial highlighted the potential strengths of CAR.  Unsurprisingly, CAR schemes achieved greater between-arm covariate balance than CR.  Impressively, under RBI, SRR with a dynamic threshold, D$_A$-BCD, and PSR achieved greater power and efficiency than using CR and fitting a regression model adjusting for covariates.  If a goal is to test the marginal average treatment effect, this case study motivates using CAR with RBI over CR with a parametric test. At the same time, overall between-arm covariate balance improved following sensible CAR schemes as compared to CR.

The observed performance of SMR and benefits of CAR with RBI can be assessed in any trial data using code available in the supplement. The code is developed for continuous outcomes but can be modified for any outcome and modelling strategy.



\section{Discussion}

This work accomplishes two major purposes: (1) it extends SMR schemes to improve the quality of sequential matches, between-arm treatment covariate balance, and study/estimator efficiency and (2) it provides a case study which compares SMR with a common and a related CAR scheme and motivates adjusting for covariates through randomization rather than through a covariate-adjusted parametric model.

The case study highlights a strength of sequential matching -- the ability to randomize between two participants who are deemed similar on selected covariates. Matching schemes attempt to optimize the overall quality of matched pairs. This has an indirect effect of balancing covariates between arm. In contrast, many CAR schemes generally seek to minimize between-arm covariate imbalance without a matched pairs component. While both optimization strategies share overlapping benefits (improved between-arm covariate balance and RBI power), each may have particular value depending on the objectives of the trial.

Providing a powerful test of the marginal average treatment effect is a common primary analysis for randomized trials. Based upon this case study, the non-trivial gains in power of CAR with RBI lead one to consider what barriers remain for this strategy over CR followed by a parametric analysis. Possible major barriers include a philosophical consideration of whether the Neyman Model with a sharp-null hypothesis is appropriate and a need for easier implementation at the point of enrollment for complex CAR schemes which cannot be pre-generated. The Food and Drug Administration has approved multiple therapeutics based upon RBI which suggests, at least to some level, that it is considered a viable test of the average treatment effects.

\section{Acknowledgements and reproducible code} This work was made possible through the support and resources of: the Center for High Performance Computing at the University of Utah, the Advanced Computing Center for Research and Education at Vanderbilt University, and the grant NIDDK R01 DK100694. We thank the anonymous reviewers for their thought-provoking and methodologically-advancing feedback.

All code for this paper is provided in the supplement and can be used to assess the benefits of SMR and CAR plus RBI using any clinical trial data with continuous outcomes. More broadly, any outcome data and modelling can be used with making minor modifications as highlighted in the README file.  Deidentified REACH study data is available upon request by contacting the principal investigator at Lindsay.Mayberry@vumc.org.  

\section{Bibliography}
\bibliographystyle{unsrt}
\bibliography{2023_07_refs}%

\end{document}